\newif\ifacm
\acmfalse
\ifacm
\documentclass[review=true]{acmart}
\else
\documentclass{amsart}
\fi

\usepackage[all]{xy}
\usepackage{bussproofs}
\usepackage{listings}
\ifacm \else
\usepackage[utf8]{inputenc}
\usepackage{stmaryrd}
\usepackage{qsymbols}
\usepackage{hyperref}
\usepackage{newunicodechar}
\DeclareUnicodeCharacter{2192}{\ensuremath{\to}}
\DeclareUnicodeCharacter{2299}{\ensuremath{\odot}} 
\DeclareUnicodeCharacter{2B3}{r} 
\fi

\usepackage[shortcuts]{extdash}

\usepackage{color}
\definecolor{darkGreen}{rgb}{0,.5,0}
\definecolor{mauve}{rgb}{1,0,1}
\definecolor{rougefonce}{cmyk}{.3,1,.3,0}
\definecolor{cyanp}{cmyk}{.5,.3,0,0}
\definecolor{yellow}{cmyk}{0,0,.7,0}
\definecolor{beige}{cmyk}{0,.2,.7,0}
\definecolor{brun}{cmyk}{0,.5,.7,0}
\definecolor{darkBrown}{cmyk}{.3,.75,.75,.15}

\newcommand{\brunir}[1]{{\color{darkBrown} #1}}

  {\color{blue}}
  {}
  {\color{darkBrown}}
  {}

  {\color{red}}
  {} %
\newtheorem{definition}{Definition}
\newcommand{\val}[1]{\ensuremath{\llbracket #1\rrbracket_{\rho}}}
\newcommand{\W}{\mathcal{W}}
\newcommand{\node}[1]{*++[o][F]{\scriptstyle #1}}
\newcommand{\noeud}{\to \ar@{-}[dl]\ar@{-}[dr]}
\newcommand{\lf}[1]{*++[F]{\scriptstyle #1}}

\newcommand{\Keywords}
{intuitionistic logic,
  classical logic,
  combinatorics,
  asymptotic,
  random generation,
  Bell number,
  Catalan number,
  Monte-Carlo method
  }

\begin{document}

\ifacm
\else

\title{Almost all classical theorems are intuitionistic}
\author{Pierre Lescanne}
\address{École Normale Sup\'erieure de Lyon,\\
  LIP (UMR 5668 CNRS ENS Lyon UCBL),\\
  46 all\'ee d'Italie, 69364 Lyon, France}
  \email{pierre.lescanne@ens-lyon.fr}

\begin{abstract}
  Canonical expressions are representative of implicative propositions
  (i.e., propositions with only implications) up-to renaming of variables.
  Using a Monte-Carlo approach, we explore the model of canonical
  expressions in order to confirm the paradox that says that
  \emph{asymptotically almost all classical theorems are intuitionistic}.
  Actually we found that more than $96,6\%$ of classical theorems are
  intuitionistic among propositions of size $100$.

\textbf{AMS Classification numbers:} 11B73, 03F55, 06E30, 05-04, 05-08

\end{abstract}
\maketitle

\fi
\section{Introduction}
\label{sec:intro}

In 2007, Marek Zaionc coauthored two
papers~\cite{DBLP:conf/types/GenitriniKZ07,DBLP:conf/csl/FournierGGZ07},
corresponding to two models of the calculus of implicative propositions
and presenting the following paradox, namely that \emph{asymptotically
  almost all classical theorems are intuitionistic}, which we call, in
short, \emph{Zaionc paradox}.  This says that when the size of the
propositions grows, the ratio of the number of intuitionistic theorems
over the number of classical theorems goes up to one.  In the current
paper, we focus on the model of~\cite{DBLP:conf/types/GenitriniKZ07},
which we call \emph{canonical expressions}.  They have been introduced by
Genitrini, Kozik and Zaionc ~\cite{DBLP:conf/types/GenitriniKZ07} and more
recently by Tarau and de
Paiva~\cite{DBLP:conf/lopstr/Tarau16,DBLP:journals/corr/abs-2009-10241}.
A~canonical expression is a representative of a class of implicative
propositions (propositions that contain only implication $"->"$) which differ only by the name assigned to the
variables. Whereas Genitrini, Kozik and Zaionc addressed the mathematical
aspect of this model, Tarau and de Paiva tried to explicitly generate all
the canonical expressions of a given size and faced up to combinatorial
explosion, because canonical expressions grow super exponentially in size when the number on variables increases.
In this paper, I check experimentally Zaionc paradox, adopting a
Monte-Carlo approach to observe how this paradox emerges.  Indeed I
designed a linear algorithm to randomly generate canonical expressions.
Therefore I can consider large samples of random (for a uniform
distribution) canonical expressions and count how many canonical
expressions in that samples are intuitionistic theorems or classical
theorems.  The experiments, centered around canonical expressions of size
$100$, show that the numbers we get for both sets are very close
confirming experimentally the paradox, with a ratio $96.6\%$ much better
than this obtained on the model of Genitrini et al. which yields $36\%$
for canonical expressions of size $100$.  As a by product we obtain
programs generating large random canonical expressions, large random
intuitionistic theorems or large random classical theorems.

The programs used in this paper can be found on
\href{https://github.com/PierreLescanne/CanonicalExpression}{\textsf{GitHub}}.

\section{Intuitionistic vs classical theorems}
\label{sec:IvsC}

In this paper we deal only with implicative propositions.  An implicative
proposition is a binary expression with propositional variables, which has
only one binary operator namely the implication written $\to$.  This can
be seen as the type of a function in functional programming or in
$\lambda$-calculus~\cite{DBLP:books/daglib/0032840,mimram:pp}.  Among the implicative propositions, some can be proven,
using a proof system, the so called \emph{natural deduction}~\cite{gentzen35}.  Let us consider natural deduction.  There are three
rules used to prove \emph{intuitionistic theorems}.

\begin{center}
  \LeftLabel{\textsf{Axiom}}
\AxiomC{\phantom{$\vdash $}}
\UnaryInfC{$\alpha \vdash  \alpha$}
\DisplayProof
\qquad
\LeftLabel{$\to$\textsf{-Elim}}
\AxiomC{$\Gamma\vdash \alpha \to \beta$}
\AxiomC{$\Gamma\vdash \alpha$}
\BinaryInfC{$\Gamma\vdash \beta$}
\DisplayProof
\qquad
\LeftLabel{$\to$\textsf{-Intro}}
\AxiomC{$\alpha,\Gamma\vdash \beta$}
\UnaryInfC{$\Gamma\vdash \alpha\to\beta$}
\DisplayProof
\end{center}

\emph{Classical theorems} are proved by adding the axiom:
\begin{prooftree}
  \LeftLabel{\textsf{Pierce}}
  \AxiomC{~}
  \UnaryInfC{$\vdash  ((\alpha \to\beta) \to\alpha) \to\alpha$}
\end{prooftree}
which is called \emph{Peirce law}.  Usually one uses valuations, which
assign booleans to variables.  Let $\rho$ be an assignment of booleans to
variables. Valuations of expressions are defined by:
\begin{eqnarray*}
  \val{x} &=& \rho(x)\\
  \val{e \to e'} &=& \val{e'} \vee \overline{\val{e}}
\end{eqnarray*}
where $b \mapsto \overline{b}$ is the negation and $b_1 \vee b_2 = 1$ except
when $b_1 = b_2 = 0$ is \emph{or}.  An expression $e$ is a classical theorem or a
tautology, if for all valuation $\rho$, $\val{e} = \mathsf{True}$.

Notice that, with the Curry Howard isomorphism~\cite{DBLP:books/daglib/0032840,mimram:pp}, the
results of this paper apply also to types.

\section{The model of canonical expressions}
\label{sec:CanExp}

We call \emph{canonical expression} the representative of an equivalence
class of binary expressions up-to renaming of variables.  In other words,
a canonical expression can be seen as a binary expression, in which variables are
named canonically, from right to left.  That means that the rightmost
variable is $x_0$, then if processing to the left, the next new variable
is $x_1$, then the next new variable, which is neither $x_0$ nor $x_1$ is
$x_2$ etc.  Recall that in an expression, a variable corresponds to a
position into the expression. In other words a variable in an equivalence
class of positions.  Therefore naming canonically a variable corresponds
to naming canonically an equivalence class in the set of position.
Therefore if a variable belongs to the $i^{th}$ class it will be named
$\alpha_i$ and vice-versa, if a class is the class of $\alpha_i$, it is the
$i^{th}$ class.  In canonical expressions, the classes are numbered from
right to left.  For instance, assume an expression of size~$10$, i.e.,
with $10$ occurrences of variables.  This is an expression with $10$
positions of variables, like :
\begin{displaymath}
x ~~ y ~~ y  ~~ x~~y ~~x ~~ z ~~ x ~~ x ~~ x
\end{displaymath}
or
\begin{displaymath}
  \beta ~~ \alpha ~~ \alpha ~~ \beta ~~ \alpha ~~ \beta ~~ \gamma ~~\beta ~~ \beta ~~ \beta
\end{displaymath}
In the first expression we see $3$ variables namely $\{x, y , z\}$, hence
$3$ congruence classes.  As said above, for technical reasons, not hard to
guess, variables are numbered from right to left, starting at $0$.  Hence
$x$ which corresponds to positions $\{1, 4, 6, 8, 9, 10\}$ is class $0$,
$z$ which corresponds to positions $\{7\}$ is class $1$ and $y$ which
corresponds to positions $\{2, 3, 5 \}$ is class $2$.  Therefore the list
of canonically named variables, associated with the above list of variables
is.
\begin{displaymath}
\alpha_0~\,\alpha_2~\,\alpha_2~\,\alpha_0~\,\alpha_2~\,\alpha_0~\,\alpha_1~\,\alpha_0~\,\alpha_0~\,\alpha_0
\end{displaymath}
The above congruence class is canonically represented by the
string $0220201000$.   Canonical presentation of congruence
classes over a set of $n$ elements by a string of natural numbers of size
$n$ is known.  It is called \emph{restricted growth string} by Knuth
(\cite{KnuthVol4_3}, fascicle 3, §7.2.1.5, p. 62) and \emph{irregular
  staircase} by Flajolet and Sedgewick~\cite{flajolet08:_analy_combin}
(p. 62-63).  In this paper, we consider classes from right to left wherever
the cited authors consider them from left to right, but this is a detail.
Intuitively, a restricted growth string is a string whose last item is $0$
and when one progresses to the left, one meets items that have been met
already or if not the new item is just the successor of the largest item
met until this point.  Here $[0..i]^*$ is the set of strings made
of integers $k$ such that $0\le k \le i$.
\begin{definition}[Restricted growth string]
  The set $\W_n$ of $n$-\emph{restricted right to left growth strings} is defined as follows
  \begin{itemize}
  \item $\W_0 = [0..0]^*$,
  \item $\W_{n+1} = [0..(n+1)]^*\, (n+1)\, \W_n$ 
  \end{itemize}
\end{definition}
For instance
$W_2 = [0..2]^*\, 2\, \W_1 = [0..2]^* \,2\, [0..1]^*\, 1\, \W_0 = [0..2]^*
\,2\, [0..1]^*\, 1\, [0..0]^*$ One sees that
$0220\brunir{2}0\brunir{1}00\brunir{0} \in \W_2$ where items larger that those
on the right are put in brown.

Once the variables are chosen, how operators~$\to$ are associated has to be
done.  Here we are interested in parenthesized expressions with the only
binary operator~$\to$.  For instance, for an expression of size $10$, we look
for a binary tree with $10$ external leaves like:
\begin{displaymath}
  \begin{scriptstyle}
 ((\Box\to\Box)\to((((\Box\to\Box)\to\Box)\to\Box)\to(((\Box\to\Box)\to\Box)\to\Box)))
\end{scriptstyle}
\end{displaymath}
which can be drawn as the tree : 
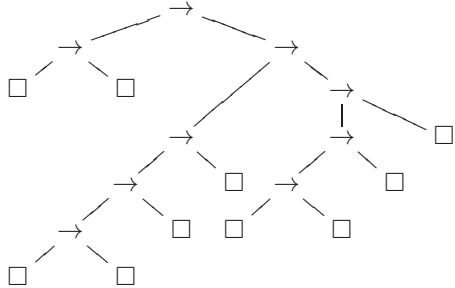
\begin{displaymath}
  \xymatrix @C=5pt @R=5pt{
    &&&\to\ar@{-}[dll]\ar@{-}[drr] \\
    & \to\ar@{-}[dl]\ar@{-}[dr] &&&& \to\ar@{-}[ddll]\ar@{-}[dr]\\
    \Box&&\Box &&&& \to\ar@{-}[d]\ar@{-}[drr]\\
    &&&\to\ar@{-}[dl]\ar@{-}[dr]&&&\to\ar@{-}[dl]\ar@{-}[dr]&&\Box\\
    &&\to\ar@{-}[dl]\ar@{-}[dr]&&\Box &\to\ar@{-}[dl]\ar@{-}[dr]&&\Box \\
    &\to\ar@{-}[dl]\ar@{-}[dr] &&\Box&\Box&&\Box\\
    \Box &&\Box 
    }
  \end{displaymath}
  To get a canonical expression one matches a restricted growth string and
  a binary tree.  In our case, we get by matching the above restricted
  tree and the above parenthesized expression, the following canonical
  expression
  \begin{displaymath}
    ((\alpha_0 \to\alpha_2 )\to((((\alpha_2 \to\alpha_0 )\to\alpha_2 )\to\alpha_0 )\to(((\alpha_1 \to\alpha_0 )\to\alpha_0 )\to\alpha_0 )))
  \end{displaymath}
  which corresponds to the tree:
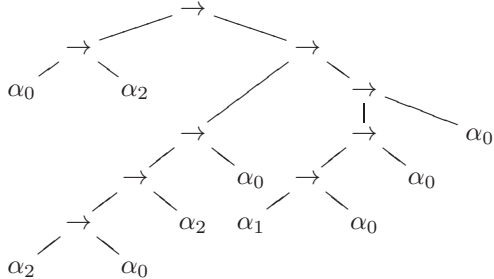
\begin{displaymath}
  \xymatrix @C=5pt @R=5pt{
    &&&\to\ar@{-}[dll]\ar@{-}[drr] \\
    & \to\ar@{-}[dl]\ar@{-}[dr] &&&& \to\ar@{-}[ddll]\ar@{-}[dr]\\
    \alpha_0&&\alpha_2 &&&& \to\ar@{-}[d]\ar@{-}[drr]\\
    &&&\to\ar@{-}[dl]\ar@{-}[dr]&&&\to\ar@{-}[dl]\ar@{-}[dr]&&\alpha_0\\
    &&\to\ar@{-}[dl]\ar@{-}[dr]&&\alpha_0 &\to\ar@{-}[dl]\ar@{-}[dr]&&\alpha_0 \\
    &\to\ar@{-}[dl]\ar@{-}[dr] &&\alpha_2&\alpha_1&&\alpha_0\\
    \alpha_2 &&\alpha_0 
    }
  \end{displaymath}

  Canonical expressions are therefore pairs of binary trees and restricted
  left to right growth strings, counted by $K_n = C_{n-1} \varpi_n$ where
  $C_n$ are \emph{Catalan numbers} (counting binary trees) and $\varpi_n$ are
  \emph{Bell numbers} (counting restricted growth strings).  This
  corresponds to sequence \href{https://oeis.org/A289679}{A289679} in the
  \emph{Online encyclopedia of integer sequences}~\cite{sloane:_OEIS}.  Asymptotically,
  \begin{displaymath}
    C_{n-1} \sim \frac{4^{n-1}}{\sqrt{\pi (n-1)^3}}
  \end{displaymath}
  \begin{displaymath}
    \varpi_n \sim n! \frac{e^{e^r -1}}{r^n \sqrt{2\pi r (r+1)e^r}}
  \end{displaymath}
  where $r \equiv r(n)$  is the positive root of the equation $r e^r = n+1$.     Therefore
  \begin{displaymath}
    K_n \quad \sim \quad  n! \frac{4^{n-1} e^{e^r -1}}{\pi \sqrt{2 (n-1)^3 r(r+1) e^r}}
  \end{displaymath}

  The first values of $K_n$ are
  $1,2,10,75,728,8526,115764,1776060,30240210,...$\\
whereas ${K_{100} \sim 9.62~ 10^{168}}$ and $K_{400} \sim 1.51~10^{880}$.

\section{Random canonical expressions}
\label{sec:Rand}
Since canonical expressions are pairs of well-known combinatorial objects,
namely binary trees and congruence classes, we can use well-known algorithm
to generate each constituents of the pairs.

\subsection{Random binary trees}
\label{sec:RandBinTree}

For generating random binary trees, I use \emph{R\'emy
  algorithm}~\cite{DBLP:journals/ita/Remy85} which is linear.  This
algorithm is described by Knuth in~\cite{KnuthVol4_3} § 7.2.1.6
(pp.~18-19).  I have taken his implementation.  The idea of the algorithm
is that a random binary tree can be built by iteratively and randomly
picking an internal node or a leaf in a random binary tree and inserting
a new internal node and a new leaf either on the left or on the right.  A
binary tree of size $n$ has $n-1$ internal nodes and $n$ leaves.
Inserting a node in a binary tree of size $n$ requires throwing randomly a
number between $1$ and $4n-2$ (a random number between $0$ and $4n-3$ in my
\textsf{Haskell} implementation).  This process can be optimized by
representing a binary tree as a list (a vector in \textsf{Haskell}), an
idea sketched by R\'emy and described by Knuth.  In this vector, even
locations are for internal nodes and odd locations are for leaves.  Here
is a vector representing a binary tree with 10 leaves and its drawing.

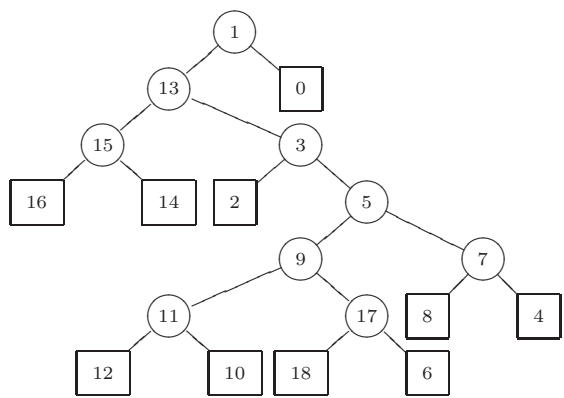
\begin{displaymath}
\begin{array}{|l||r|r|r|r|r|r|r|r|r|r|r|r|r|r|r|r|r|r|r|r|r|r|}
  \hline
  \textbf{\footnotesize indices}&0&1&2&3&4&5&6&7&8&9&10&11&12&13&14&15&16&17&18\\
  \hline
  \textbf{\footnotesize values}&1&13&0&2&5&9&7&8&4&11&17&12&10&15&3&16&14&18&6\\
  \hline
\end{array}
\end{displaymath}
\begin{displaymath}
  \xymatrix @C=5pt @R=5pt{
    &&&\node{1}\ar@{-}[dl]\ar@{-}[dr]\\ 
    &&\node{13}\ar@{-}[dl]\ar@{-}[drr]&&\lf{0}\\ 
    &\node{15}\ar@{-}[dl]\ar@{-}[dr]&&&\node{3}\ar@{-}[dl]\ar@{-}[dr]\\ 
    \lf{16}&&\lf{14}&\lf{2}&&\node{5}\ar@{-}[dl]\ar@{-}[drr]\\ 
    &&&&\node{9}\ar@{-}[dll]\ar@{-}[dr]&&&\node{7} \ar@{-}[dl]\ar@{-}[dr]&\\ 
    &&\node{11}\ar@{-}[dl]\ar@{-}[dr] &&& \node{17}\ar@{-}[dl]\ar@{-}[dr] &\lf{8}&&\lf{4}\\ 
    &\lf{12 }&& \lf{10} & \lf{18}&&\lf{6}
    }
  \end{displaymath}

  This tree was built by inserting the node $17$ together with the leaf $18$ in the following tree. 
  \begin{displaymath}
    \begin{array}{|l||r|r|r|r|r|r|r|r|r|r|r|r|r|r|r|r|r|r|r|r|}
  \hline
  \textbf{\footnotesize indices}&0&1&2&3&4&5&6&7&8&9&10&11&12&13&14&15&16\\
  \hline
  \textbf{\footnotesize values}&1&13&0&2&5&9&7&8&4&11&6&12&10&15&3&16&14\\
  \hline
\end{array}
  \end{displaymath}
  coding the tree
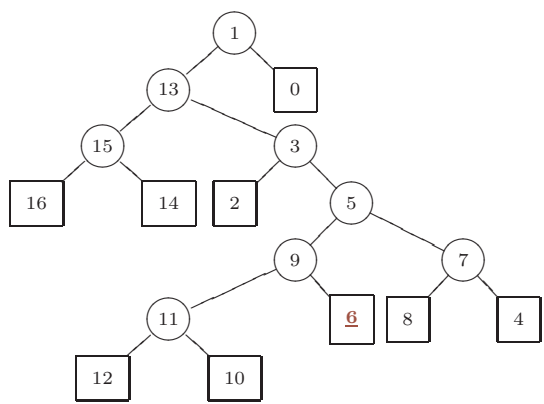
\begin{displaymath}
  \xymatrix @C=5pt @R=5pt{
    &&&\node{1}\ar@{-}[dl]\ar@{-}[dr]\\ 
    &&\node{13}\ar@{-}[dl]\ar@{-}[drr]&&\lf{0}\\ 
    &\node{15}\ar@{-}[dl]\ar@{-}[dr]&&&\node{3}\ar@{-}[dl]\ar@{-}[dr]\\ 
    \lf{16}&&\lf{14}&\lf{2}&&\node{5}\ar@{-}[dl]\ar@{-}[drr]\\ 
    &&&&\node{9}\ar@{-}[dll]\ar@{-}[dr]&&&\node{7} \ar@{-}[dl]\ar@{-}[dr]&\\ 
    &&\node{11}\ar@{-}[dl]\ar@{-}[dr] &&& *++[F]{\scriptstyle \brunir{\textbf{\underline{6}}}} &\lf{8}&&\lf{4}\\ 
    &\lf{12 }&& \lf{10}
    }
  \end{displaymath}
  This was done by picking a node (internal node or leaf, here the node
  with label $6$) and a direction (here right) and by inserting above this
  node a new internal node (labeled $17$) and, below the new inserted
  internal node, a new leaf of the left (labeled $18$).  This double action (inserting the internal
  node and attaching the leaf) is done by choosing a number in the
  interval $[0..33]$ (in general, in the interval $[0..(4n-3)]$). Assume
  that in this case the random generator returns~$21$.  $21$ contains two
  informations : its parity (a boolean) and its half.  Half of $21$ is
  $10$, which tells that the new node $17$ must be inserted above the
  $10^{th}$ (in the array) node namely~$6$. Since $21$ is odd, the rest of
  the tree (here reduced to the leaf $6$) is inserted on the right
  (otherwise it would be inserted on the left).  A new leaf $18$ is
  inserted on the left (otherwise it would be inserted on the right).

  Consider same tree and suppose that the random value is $8$. Half of $8$ is $4$. Hence the new leaves are inserted above the node labeled by $5$
  
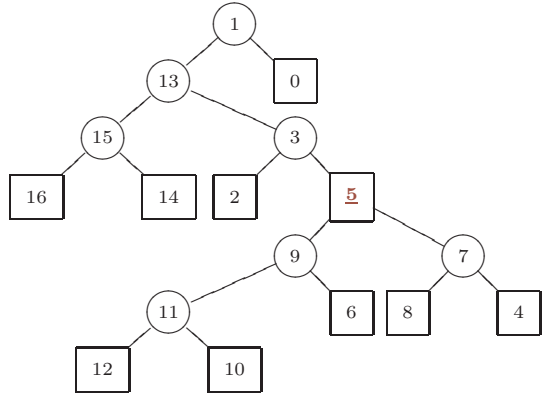
\begin{displaymath}
  \xymatrix @C=5pt @R=5pt{
    &&&\node{1}\ar@{-}[dl]\ar@{-}[dr]\\ 
    &&\node{13}\ar@{-}[dl]\ar@{-}[drr]&&\lf{0}\\ 
    &\node{15}\ar@{-}[dl]\ar@{-}[dr]&&&\node{3}\ar@{-}[dl]\ar@{-}[dr]\\ 
    \lf{16}&&\lf{14}&\lf{2}&&*++[F]{\scriptstyle \brunir{\textbf{\underline{5}}}}\ar@{-}[dl]\ar@{-}[drr]\\ 
    &&&&\node{9}\ar@{-}[dll]\ar@{-}[dr]&&&\node{7} \ar@{-}[dl]\ar@{-}[dr]&\\ 
    &&\node{11}\ar@{-}[dl]\ar@{-}[dr] &&& \lf{6} &\lf{8}&&\lf{4}\\ 
    &\lf{12 }&& \lf{10}
    }
  \end{displaymath}
  and since $8$ is even the rest of tree is inserted on the left and a new leaf (labeled $18$) is inserted on the right. 
  
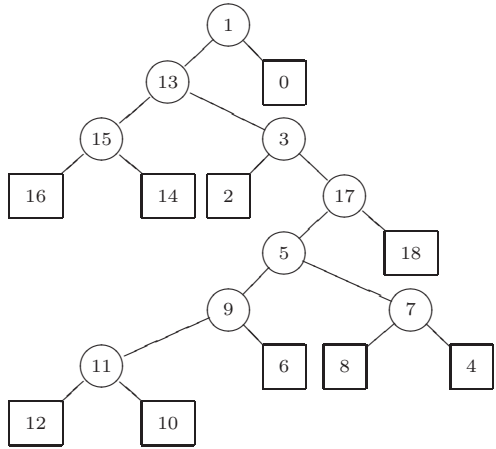
\begin{displaymath}
  \xymatrix @C=5pt @R=5pt{
    &&&\node{1}\ar@{-}[dl]\ar@{-}[dr]\\ 
    &&\node{13}\ar@{-}[dl]\ar@{-}[drr]&&\lf{0}\\ 
    &\node{15}\ar@{-}[dl]\ar@{-}[dr]&&&\node{3}\ar@{-}[dl]\ar@{-}[dr]\\ 
    \lf{16}&&\lf{14}&\lf{2}&&\node{17}\ar@{-}[dl]\ar@{-}[dr]\\
    &&&&\node{5}\ar@{-}[dl]\ar@{-}[drr]&&\lf{18}\\ 
    &&&\node{9}\ar@{-}[dll]\ar@{-}[dr]&&&\node{7} \ar@{-}[dl]\ar@{-}[dr]&\\ 
    &\node{11}\ar@{-}[dl]\ar@{-}[dr] &&& \lf{6} &\lf{8}&&\lf{4}\\ 
    \lf{12 }&& \lf{10}
    }
  \end{displaymath}

  The algorithm works as follows. If $n=0$, R\'emy's algorithm returns the
  vector starting at $0$ and filled with anything, since the whole
  algorithm works on the same vector with the same size.  In general, say
  that, for $n-1$, R\'emy's algorithm returns a vector $v$.  One picks a
  random integer $x$ between $0$ and $4n-3$. Let $k$ be half of $x$.  In
  the vector $v$ one replaces the $k^{th}$ position by $2n-1$ and one
  appends two elements, namely the $k^{th}$ item of $v$ followed by $2n$
  if $x$ is even and $2n$ followed by the $k^{th}$ item of $v$ if $x$
  is odd.

  If we admit that given a \textsf{seed} and a positive integer
  \textsf{n}, \texttt{randForRemy seed n } returns a random integer
  between $0$ and $4 n - 3$ inclusive, the program in \textsf{Haskell} of
  the function \texttt{rbtV} which yields a random binary tree of size $n$
  coded as a vector of length $2n$ is given in Fig.~\ref{fig:program}.
  \begin{figure}[!th]
    \centering
\begin{lstlisting}
rbtV :: Int -> Int -> Vector Int
rbtV seed 0 = Data.Vector.replicate sizeOfVector (-1) // [(0,0)]
rbtV seed n =
  let x = randForRemy seed n -- a random value between 0 and 4n-3 inclusive
      v = rbtV seed (n-1)
      k = x `div` 2
  in case even x of
    True -> v // [(k,2*n-1),(2*n-1,v!k),(2*n,2*n)]
    False -> v // [(k,2*n-1),(2*n-1,2*n),(2*n,v!k)]
\end{lstlisting}

\caption{Haskell program for R\'emy's algorithm}
\label{fig:program}
\end{figure}

\subsection{Random restricted growth string}
\label{sec:RandStr}
For generating random partitions or random restricted growth strings an
algorithm due to A.~J.~Stam~\cite{DBLP:journals/jct/Stam83} and described
by Knuth in~\cite{KnuthVol4_4} § 7.2.1.3 (p.~74) was implemented.  The
implementation requires, for each value of $n$ (the size of the underlying
set --\,for us, this is the number of variables or the size of the expression\,--), a
preliminary construction of a table of reals in which indices are looked
up (the number $M$ of classes). Those reals are probabilities
\begin{displaymath}
  p_n = \frac{m^n}{e m! \varpi_n}
\end{displaymath}
that an $n$-partition has $m$ classes.  Thus in my program, I implemented
the algorithm for size $n = 10$, $25$, $50$ $100$, $500$ and $1000$.  In
order to get accurate values, the $p_n$'s for those integers were
computed elsewhere in a dedicated computer algebra software namely
\textsf{Sagemath}~\cite{sagemath}.  From this table and a randomly chosen
number between $0$ and $1$, one gets a random number $M$ of equivalence
classes. Thereafter, for each element in $[0..n]$ one picks up randomly
uniformly and independently individuals in $[0..(M-1)]$.  This method
yields \emph{class descriptions} (classes are a priori numbered from $0$
to $M-1$ and the elements $0$,..., $n-1$ are distributed in those
classes), but one wants \emph{restricted growth strings} as described in
Section~\ref{sec:CanExp}.  So a function that transforms a class
description into a restricted growth string was implemented.

Putting together those two algorithms, namely binary tree random
generation and restricted growth string random generation, produces an
algorithm for canonical expression random generation.

\section{Selecting intuitionistic theorems}
\label{sec:Int}

Once a canonical expression is randomly generated, one has to check
whether it is an intuitionistic theorem, a classical theorem, or not a
proposition of those sorts

The program selects two kinds of trivial intuitionistic expressions.  At
first glance this selection looks coarse, but from experience, the first
one (\emph{simple} theorems) collects a large majority of the expressions
and the second selects (\emph{arrowElim} theorems) most of the others,
because it is associated with a trick which consists in cleaning
expressions by removing recursively trivial subexpressions that are
theorems.  Indeed a ``cleaned'' sub-expression can become trivial and be
removed in turn.  This might allow cleaning an expression where a trivial
premise appears, which might be removed in turn.  Therefore this section
lists six methods for selecting more and more intuitionistic theorems.
Except ``simple'', these adjectives are mine.

\subsection{Simple intuitionistic theorems}
\label{sec:simple}
A \emph{simple intuitionistic theorem} (see~\cite{DBLP:conf/types/GenitriniKZ07} Definition 1) is a theorem, in which the
goal is among the premises.  In other words, this is a theorem of the
form:
\begin{displaymath}
  ... \to \alpha_i \to ... \to \alpha_i
\end{displaymath}

\subsection{MP intuitionistic theorems}
\label{sec:elim}

Let us call \emph{MP intuitionistic theorem} (for \emph{modus ponens} theorem), a theorem which is a direct
application of the modus ponens.  This is a theorem with goal $\alpha_i$ and two
premises $\alpha_j$ and $\alpha_j\to \alpha_i$.  Therefore it has the form:
\begin{displaymath}
  ... \to (\alpha_j\to\alpha_i)\to ... \to\alpha_j\to ... \to \alpha_i
\end{displaymath}
or
\begin{displaymath}
  ... \to\alpha_j\to ... \to (\alpha_j\to\alpha_i)\to ... \to \alpha_i
\end{displaymath}

During the experiments I met, for instance, the term, which is not a
canonical expression, but obtained by cleaning:
\begin{scriptsize}
  \begin{displaymath}
(x_{28}\to((x_{22}\to((x_{26}\to((x_{14}\to x_{2})\to(x_{11}\to x_{8})))\to x_{28}))\to((x_{28}\to(x_{9}\to x_{13}))\to(x_{14}\to((x_{28}\to x_{0)}\to x_{0})))))  
\end{displaymath}
\end{scriptsize}
which can be drawn as the labeled binary tree:
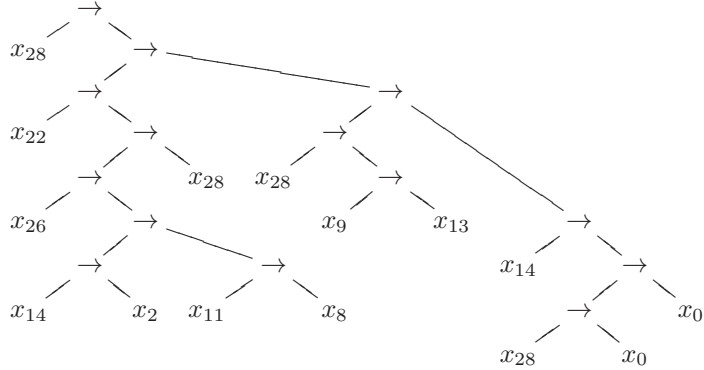
\begin{displaymath}
  \xymatrix @C =5pt @R = 5pt{
    & \noeud\\ 
    x_{28} &&\to \ar@{-}[dl]\ar@{-}[drrrr]\\ 
    &\noeud&&&&&\to \ar@{-}[dl]\ar@{-}[dddrrr] \\ 
    x_{22} &&\noeud &&&\noeud&&&&\\ 
    &\noeud&&x_{28}&x_{28}&&\noeud\\ 
    x_{26}&&\to \ar@{-}[dl]\ar@{-}[drr]&&&x_9&&x_{13}&&\noeud\\ 
    &\noeud&&& \to \ar@{-}[dl]\ar@{-}[dr]&&&&x_{14}&&\noeud&&\\ 
    x_{14}&&x_{2}&x_{11}&&x_{8} &&&&\noeud&&x_0 \\
    &&&&&&&&x_{28}&&x_0
    }
\end{displaymath}
which can be written
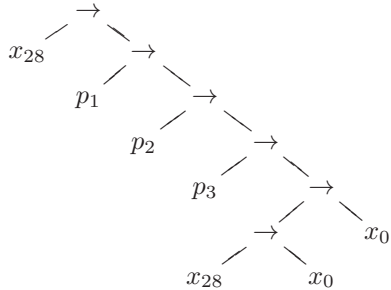
\begin{displaymath}
  \xymatrix  @C =5pt @R = 5pt{
    &\noeud \\ 
     x_{28} && \noeud \\ 
    & p_1 && \noeud \\ 
    && p_2 && \noeud \\ 
    &&& p_3 && \noeud \\ 
    &&&& \noeud && x_0 \\ 
    &&& x_{28} && x _0 
    }
  \end{displaymath}
  It is clearly an intuitionistic theorem and \textsf{isMP} checks it.

\subsection{Easy intuitionistic theorems}
\label{sec:easy}

Let us call \emph{easy intuitionistic theorems}, expressions that are
\emph{simple} or \emph{mp}.

\subsection{Removing easy premises}
\label{sec:triv}

In intuitionistic logic if a premise is a theorem, it can be
removed. Consider the predicate $\vdash p$ that says that $p$ is a theorem.
Clearly under the assumption $\vdash p$, the two statements $\vdash  p \to q$ and
$\vdash q$ are equivalent.  Note that $p$ is not necessarily the first
premise of the implication.  Hence if an expression becomes \emph{easy
  after removing easy premises}, it is an intuitionistic theorem.

In the process of ``cleaning'' expressions, expressions that are \emph{easy} are
removed inside-out.  This way \emph{easy} expressions that can be removed
recursively are detected.

\subsection{Minor intuitionistic theorems}
\label{sec:minor}

A \emph{minor} theorem is a theorem of the form
$ ... \to p \to ... \to p$, whatever $p$ is. \emph{Simple} propositions
are \emph{minor}, but \emph{minor} propositions are not always \emph{simple}.
For instance, $x\to(y\to z)\to y \to z$ is minor, but is not
\emph{simple}.  Detecting such expressions has a cost, I decided to not
detect \emph{minor} intuitionistic theorem recursively, but only after
\emph{easy} subexpressions have been removed recursively.

\subsection{Cheap intuitionistic theorems}
\label{sec:cheap}

Let us call \emph{cheap intuitionistic theorems}, expressions that are
\emph{minor} or \emph{easy} after removing (recursively) \emph{easy}
premises.  Actually, my experiments lead naturally to the statement that
96.6\% of classical theorems with $100$ variables are \emph{cheap}
intuitionistic (see Section~\ref{sec:results}).

\section{Classical tautologies}
\label{sec:taut}
The selection of classical tautologies is by valuations.  If all the
valuations of an expression yield \textsf{True} this expression is a
classical tautology.  But this method is obviously
intractable~\cite{DBLP:conf/stoc/Cook71}.  It should be applied only to
expressions on which other more efficient methods do not work and with a
limitation on the number of variables in expressions

\subsection{Simple antilogies}
\label{sec:my}

Trivial non classical propositional theorems are eliminated before
applying valuations.  The predicate \textsf{simpAntilogy} finds in
quadratic time a large set of propositions which are not tautologies and
which we call \emph{simple antilogies}.  Thereafter, boolean valuations are
checked only on the positions that are not \emph{simple antilogies}.  For more efficiency, the
predicate \textsf{simpAntilogy} is applied on expressions in which easy
premises have been recursively removed, like for intuitionistic
expressions.

An expression is a \emph{simple antilogy} if it is of the form
\begin{math}
  ... \to e_i \to ... \to x_0
\end{math}
where the premises $e_i$ are of one of the following forms:
\begin{itemize}
\item[(i)] $\to ... \to x_i$ \quad with $x_i \neq x_0$ \quad i.e., with a goal which is not $x_0$
\item[(ii)] \begin{math}
  ... \to x_0 \to ... \to x_0
\end{math}
\quad i.e., are \emph{simple} with goal $x_0$.
\end{itemize}
One sees easily that applying the valuation $\rho$ such that
$\rho(x_0) = \mathsf{False}$ and $\rho(x_i) = \mathsf{True}$ for $i\neq 0$
to \emph{simple antilogies} yields $False$. Therefore \emph{simple antilogies} are
not classical theorems.

In~\cite{DBLP:conf/types/GenitriniKZ07}, Genitrini, Kozik and Zaionc
consider only the first case, namely the case where the premises have a
goal which is note $x_0$.  They call such expressions, \emph{simple non
  tautologies}. 

\subsection{Expressions with too many variables}
\label{sec:trick}


Assume we recursively remove \emph{simple antilogies}, there are still
expressions intractable by the valuation method, because they have too
many variables, i.e., they have a too large index.  In my experiment with
expressions of size $100$, an index is too large if it is larger than
$31$.  Fortunately those expressions are rare and one may expect that
there is a valuation that rejects them.  For this, I rename all the too
large indices as they would be the same as the bound.  The valuations are
checked on this renamed expression. If the renamed expression is not a
tautology, then the given expression is not a tautology.  In the
experiment of Section~\ref{sec:results} this trick works and eliminates
expressions with too large indices which need not to be checked further.

\section{Results}
\label{sec:results}

\subsection{Ratio cheap vs classical}
\label{sec:ratio}

My \href{https://github.com/PierreLescanne/CanonicalExpression}{Haskell
  program} was run on a sample of $20\,000$ randomly generated canonical
expressions of size $100$ and I found $759$ classical tautologies, among
which $733$ were cheap expressions, hence guaranteed to be intuitionistic
theorems.  Therefore the ratio of cheap theorems over classical theorems
is $96.6\%$.  Said otherwise, less that $3.4 \%$ of the classical theorems
are not cheap, i.e., are not intuitionistic.  Are these $3.4\%$ classical
non cheap theorems still intuitionistic?  The experience cannot tell.  I
presume that there are likely more than $733$ intuitionistic theorems and
therefore among propositions of size $100$, more than $96.6\%$ of
classical theorems that are intuitionistic, or less that $3.4\%$ of
intuitionistic classical propositions that are not inuitionistic.

\subsection{Simple intuitionistic theorems vs not simple non tautologies}
\label{sec:SvsNSNT}

In~\cite{DBLP:conf/types/GenitriniKZ07}, Genitrini, Kozik and Zaionc take
the ratio of the number of simple intuitionistic theorems over the number
of non simple non tautologies as the quantity that goes to $1$ and is a lower
bound of the ratio of the number of intuitionistic theorems over the number
of classical theorems.  Among $10\,000$ random canonical expressions of
size $100$, I found $238$ simple intuitionistic theorems and $685$ non
simple non tautologies, for a ratio closed to $36\%$, a ratio largely
smaller than the above one.

\subsection{Simple intuitionistic theorems}
\label{sec:resSimp}

Besides, another number of interest is the ratio $R_n$ of simple
intuitionistic theorems over all canonical expressions of size $n$. In the
next array, this is compared with the formula $\frac{log(n)}{n}$. 

\begin{displaymath}
  \begin{array}[h]{|r|l|l|}
    \hline
    n & \quad\frac{log(n)}{n}& R_n\\
    \hline\hline
    25  & 0,128755033 & 0.2214 \\\hline
    50  & 0,07824046 & 0.1248 \\\hline
    100 & 0,046051702 & 0.0506\\\hline 
    500 & 0,012429216 & 0.0119\\\hline 
    1000 &0,006907755 & 0.006 \\\hline 
  \end{array}
\end{displaymath}

Genitrini, Kozik and Zaionc~\cite{DBLP:conf/types/GenitriniKZ07} gave
$\frac{e\;log(n)}{n}$ in Lemma 2, for the same quantity, but after viewing
my results Genitrini~\cite{genitrini21:_errat_class_tautol_quant_compar}
found a mistake and corrected the formula to $\frac{log(n)}{n}$, which now
corresponds to what I found. Notice that this does not affect their other
results.

\subsection*{Acknowledgments}
\label{sec:ack}
I thank Valeria De Paiva for an interesting interaction and the incentive
to address this problem, Jean-Luc R\'emy for discussions on binary tree
generation and Antoine Genitrini for discussions on Zaionc paradox.

\section{Conclusion}
\label{sec:concl}

Algorithms for random generation presented in \emph{The Art of Computer
  Programming}~\cite{KnuthVol4_3,KnuthVol4_4} allow implementing
Monte-Carlo methods that confirm experimentally Zaionc paradox and show
that the convergence (as the size of the expressions grows) of
the set of intuitionistic theorems toward this of classical theorems is
faster than expected from the asymptotic approximations proposed by the
analytic combinatorial theory~\cite{DBLP:conf/types/GenitriniKZ07}.
Indeed, whereas I compare the set of cheap intuitionistic theorems
(Section~\ref{sec:cheap}) with this of classical theorems, Genitrini,
Kozik and Zaionc compare the set of simple intuitionistic theorems (see
Section~\ref{sec:simple}) with the set of non simple non tautologies
(Section~\ref{sec:my}).  This is a too rough approximation and this
suggests to complete the analytic development to justify this faster
convergence.

Notice that Tarau and de Paiva~\cite{DBLP:journals/corr/abs-2009-10241}
looked at a phenomenon similar to Zaionc paradox for linear logic.
Therefore, it should be interesting to extend my approach to this case.
Likewise, it would be interesting to investigate experimentally other
models of expressions, for both traditional logic and linear logic.
Currently I am exploring expressions made of a binary operator, like
$\wedge$, $\vee$ or $\to$.

It seems that this result on the distribution of propositions has to do
with the amazing efficiency of
SAT-solvers~\cite{DBLP:conf/mc2/BrightGKG19,biere09:_handb}. The fact that
most of the classical theorems can be solved as ``cheap''
intuitionistic propositions may explain why SAT-solvers are so efficient
and the connection should be further investigated.  Likely, the remaining
true classical propositions contribute to the hardness of SAT for the
worst case analysis.


\end{document}